# DO GAMMA-RAY BURST SOURCES REPEAT?

Charles A. Meegan[1], Dieter H. Hartmann[2], J. J. Brainerd[3], Michael S. Briggs[3], William S. Paciesas[3], Geoffrey Pendleton[3], Chryssa Kouveliotou[4], Gerald Fishman[1], George Blumenthal[5], Martin Brock[1]

[1] NASA/Marshall Space Flight Center, Huntsville, AL 35899
[2] Department of Physics and Astronomy, Clemson University, Clemson, SC 29634
[3] Department of Physics, University of Alabama in Huntsville, Huntsville, AL 35899
[4] USRA/Marshall Space Flight Center, Huntsville, AL 35812
[5] UCO/Lick Observatory, UC Santa Cruz, Santa Cruz, CA 95064

## ABSTRACT

The demonstration of repeated gamma-ray bursts from an individual source would severely constrain burst source models. Recent reports (Quashnock & Lamb 1993; Wang & Lingenfelter 1993) of evidence for repetition in the first BATSE burst catalog have generated renewed interest in this issue. Here, we analyze the angular distribution of 585 bursts of the second BATSE catalog (Meegan *et al.* 1994). We search for evidence of burst recurrence using the nearest and farthest neighbor statistic and the two-point angular correlation function. We find the data to be consistent with the hypothesis that burst sources do not repeat; however, a repeater fraction of up to about 20% of the observed bursts cannot be excluded.

*Subject headings:* gamma-rays: bursts

# 1. INTRODUCTION

The observed isotropic and inhomogeneous spatial distribution derived from BATSE burst data (Meegan *et al.* 1992; Fishman *et al.* 1994) severely constrains possible Galactic distributions and argues in favor of sources at cosmological distances. Although neutron stars in an extended Galactic halo were considered as an alternative to cosmological models, the observational constraints are now so severe that a halo origin of bursts appears unlikely (Hakkila *et al.* 1994; Hartmann *et al.* 1994b; Briggs *et al.* 1995). In contrast, cosmological models naturally explain the observed isotropy and inhomogeneity, but some of these are incapable of producing multiple bursts from one source.

The absence of an excess of overlapping error circles in pre-BATSE burst localizations provides a model-dependent upper limit of ~10 years on the burst repetition time scale (Schaefer & Cline 1985; Atteia *et al.* 1987). Several recent reports, however, have cited evidence for repetition on much shorter times. Quashnock & Lamb find an excess of close neighbors in the 1B burst catalog, and conclude that a large fraction of classical bursts repeat on timescales of order months (Quashnock & Lamb 1993). Wang & Lingenfelter (1993) claim evidence for repetition with recurrence times perhaps as short as days from one particular location (0855−00). Both Quashnock & Lamb and Wang & Lingenfelter interpret their results as evidence for multiple repetitions, of about five observed bursts per repeating source. Also, the coincidence of the locations of GRB940301 and GRB930704 determined with COMPTEL has a 3% chance probability, suggesting a possible repeater (Kippen *et al.* 1995).

To investigate burst repetition, we test the null hypothesis that "the angular distribution of GRBs is consistent with the isotropic distribution", i.e., equal probability per unit solid angle. Tests of isotropy have varying sensitivities to clustering, which can indicate the presence of repetition, as well as to large-scale anisotropies. In this paper we focus on the implications for burst repetition of the two-point correlation function and the nearest neighbor test. We do not consider possible time dependent repetition, as suggested by Wang & Lingenfelter (1993), which is discussed in a separate paper (Brainerd *et al.* 1995). Other clustering tests are considered in Hartmann *et al.* (1995). We consider only the "classical" gamma-ray bursts, which are distinct from the Soft Gamma Repeaters

| Nearest Neighbor Analysis | | | | | | | | |
|---|---|---|---|---|---|---|---|---|
| Set | Triggers | MAXBC | $> 9°$ | Size | $K_{cel}$ | $S_{cel}$ | $K_{gro}$ | $S_{gro}$ |
| 1 | 105–1466 | no | yes | 262 | 1.77 | 0.028 | -1.02 | 0.50 |
| 2 | 1467–2230 | no | yes | 223 | 0.98 | 0.54 | -1.05 | 0.51 |
| 3 | 1467–2121 | yes | yes | 262 | 1.15 | 0.34 | 1.02 | 0.49 |
| 4 | 1467–2230 | yes | yes | 323 | 0.61 | 0.98 | -1.23 | 0.27 |
| 5 | 105–2230 | no | yes | 485 | 1.41 | 0.14 | -0.54 | 0.994 |
| 6 | 105–2230 | yes | yes | 585 | 1.03 | 0.49 | -0.75 | 0.87 |
| 7 | 105–1466 | no | no | 202 | 2.66 | $1.03 \cdot 10^{-4}$ | -0.76 | 0.85 |
| 8 | 1467–2230 | no | no | 195 | 0.86 | 0.71 | -1.03 | 0.48 |

Table 1

(Kouveliotou 1994).

We analyze the 2B catalog of bursts observed by BATSE between 1991 April 19 and 1993 March 9, comprising 585 bursts (Meegan et al. 1994) and various subsets. Data after 1992 March contain numerous gaps due to CGRO tape recorder errors. For 100 bursts that were most seriously affected by these gaps, the locations are determined with MAXBC data, which consists of the background-subtracted maximum rates in each detector on a 1.024 second timescale in the 50 to 300 keV energy range. Koshut et al. (1994) find that the total location error (systematic + statistical) is about 7° for bright bursts located using MAXBC data. Since this is larger than the usual systematic error of about 4°, the inclusion of MAXBC-located bursts might obscure a repeater signal.

Our subsets, which are listed in Table 1, are of consecutive triggers limited by the listed trigger numbers. We apply two cuts to some of the data sets: data sets marked with "yes" in the MAXBC column contain bursts located with the MAXBC data, while data sets with "no" in this column have such bursts removed. Similarly, data sets with "yes" in the column "$> 9°$" contain bursts with statistical location errors $> 9°$, while data sets with "no" in this column do not. The total location error of a burst is estimated as the rms sum of its statistical error and a 4° systematic error.

We define $f$ as the fraction of all observed bursts that can be labeled as repeaters and $\nu$ as the average number of observed events per source observed to repeat (thus $\nu \geq 2$). We define $N_B$ as the number of observed bursts and $n_r$ as the number of sources from which two or more bursts were observed. These quantities are related via $n_r = fN_B/\nu$.

## 2. TWO-POINT ANGULAR CORRELATION FUNCTION

One mathematical function that is used to test for anisotropy is the two-point angular correlation function, $w(\theta)$, defined in the following manner: for an ensemble of points distributed on the sky, the average number of pairs with angular separation $\theta$ within the solid angle $d\Omega$ is (e.g., Peebles 1980)

$$dN_p = \frac{N_B - 1}{4\pi} \left[1 + w(\theta)\right] d\Omega. \quad (1)$$

The application of angular correlation analysis to GRB data was introduced by Hartmann & Blumenthal (1989) for point sources and refined for fuzzy sources by Hartmann, Linder, & Blumenthal (1991). If some of the observed bursts are repeaters, each of their positions on the sky will be displaced from the source position by a distance of order the total location error. If the location error is characterized by a Gaussian distribution with a standard deviation of $\sigma_* \ll 1$, then the the observed correlation function is

$$(N_B - 1) w(\theta) = f(\nu - 1) \left[\frac{2}{\sigma_*^2} \exp\left(-\frac{\theta^2}{2\sigma_*^2}\right) - 1\right]. \quad (2)$$

This equation shows that excess correlation is spread over an angular scale of $\sigma_*$ ($\sim 7°$ for BATSE). The negative correlation at larger angles occurs because the correlation function must integrate to zero. For a given repeater fraction $f$ the strength of a repetition signal in the data increases with $\nu - 1$.

Figure 1 shows the two-point correlation functions for data sets 1, 5 and 6 (Table 1), which are the revised 1B catalog, the 2B catalog and the 2B catalog less MAXBC-located bursts. In the 1B data two regions with excess at the $2\sigma$ level are apparent, one near $0°$, which is interpreted in Quashnock & Lamb (1993) as evidence of repetition, and one near $180°$, which Narayan & Piran (1993) use to argue against the repetition interpretation (but see Quashnock & Lamb 1994). Here we use the final localizations of the 2B BATSE catalog and find that there are no significant angular correlations on any scale in the current data set. Burst data obtained after the 1B period do not show the excesses that are apparent in the 1B set. The combined data sets still show some residual effects of the 1B excesses but the deviations are not significant. Consistency with the null hypothesis of zero correlations is evaluated with the Kuiper (1960) statistic, which has certain advantages over the usual Kolmogorov-Smirnov (KS) test in the current context (Hartmann et al. 1995). We have also applied various cuts in angular resolution, using subsets of the data with better localization accuracy. None of these sets shows evidence for significant deviations from zero clustering. To derive upper limits on the presence of observed repeaters, we fit the data with the model correlation function (eq. 2). We set $\sigma_* = 7°$, the mean positional uncertainty of BATSE locations, and fit the first two 4° wide bins of the correlation function using various values of $f(\nu - 1)$. The fits become unacceptable at the 99% confidence level if $f(\nu - 1)$ exceeds $\approx 0.2$. The most difficult case to detect is $\nu = 2$: the limit for this case is $f \lesssim 0.2$. Repetition has been previously reported for $\nu \approx 5$: the limit for this case is $f \lesssim 0.05$.

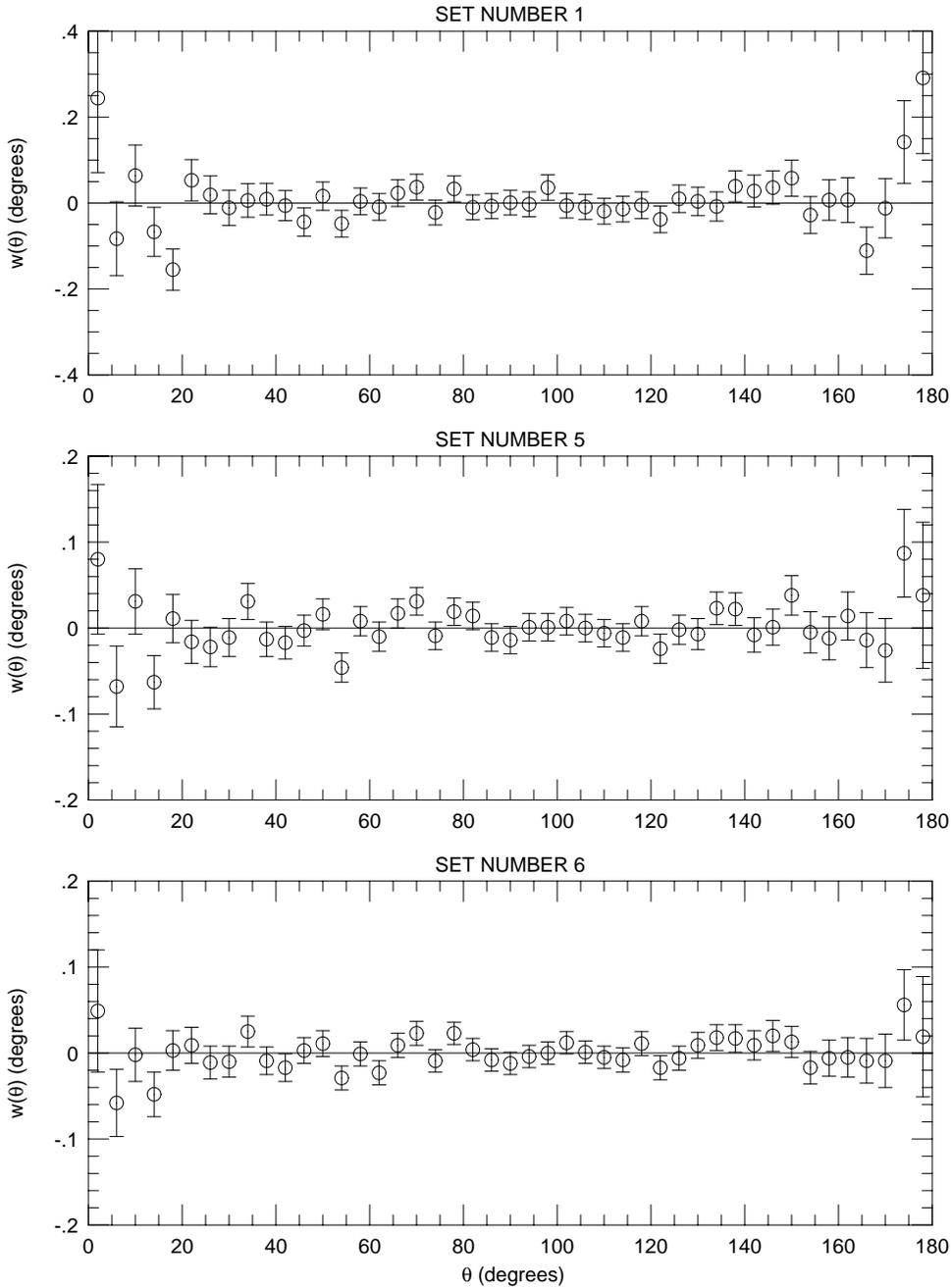

Fig. 1—The angular correlation function of gamma-ray bursts. Shown are the results for 262 bursts in the 1B catalog (data set 1), the full 2B set of 585 bursts (data set 5), and the modified 2B set with 485 bursts in which MAXBC events (see text) were removed from the sample (data set 6). The addition of second year data clearly reduces both excesses near 0° and near 180° originally found in the 1B set.

## 3. NEAREST NEIGHBOR TEST

Another approach to detecting burst recurrences is the nearest neighbor test, which tests whether the separations between bursts are consistent with the separations found for the isotropic distribution. For isotropically distributed bursts one expects the cumulative distribution of nearest neighbors to be (Scott & Tout 1989)

$$D_{\rm NN}(\theta) = 1 - \left(\frac{1 + \cos\theta}{2}\right)^{N_B - 1}. \tag{3}$$

Burst repetition will create small-scale anisotropies in the burst densities. The nearest neighbor test can indicate the existence of such anisotropies if the average distance between bursts is greater than the location error. For BATSE, this requires that the sample size be less than about 500 bursts (Brainerd et al. 1994), although multiple repetitions may be detected in larger samples.

The First BATSE Gamma-Ray Burst Catalog was analyzed by Quashnock & Lamb (1993) for burst repetition by comparing through the KS statistic the cumulative distribution of nearest neighbor separations with that expected for a uniform sky distribution. This was done for the full catalog of 260 bursts and for various subsets. They found a deviation from isotropy of 2% significance for the full catalog and of $1.1 \times 10^{-4}$ significance for the 202 bursts with statistical errors less than 9 degrees. The selection of the 9 degree error cut maximizes the signal but introduces uncertainty in the calculation of statistical significance, since the value of 9 degrees was not specified *a priori*. Such techniques are useful for exploring a data set for unanticipated effects but must be treated as predictions for subsequent data sets.

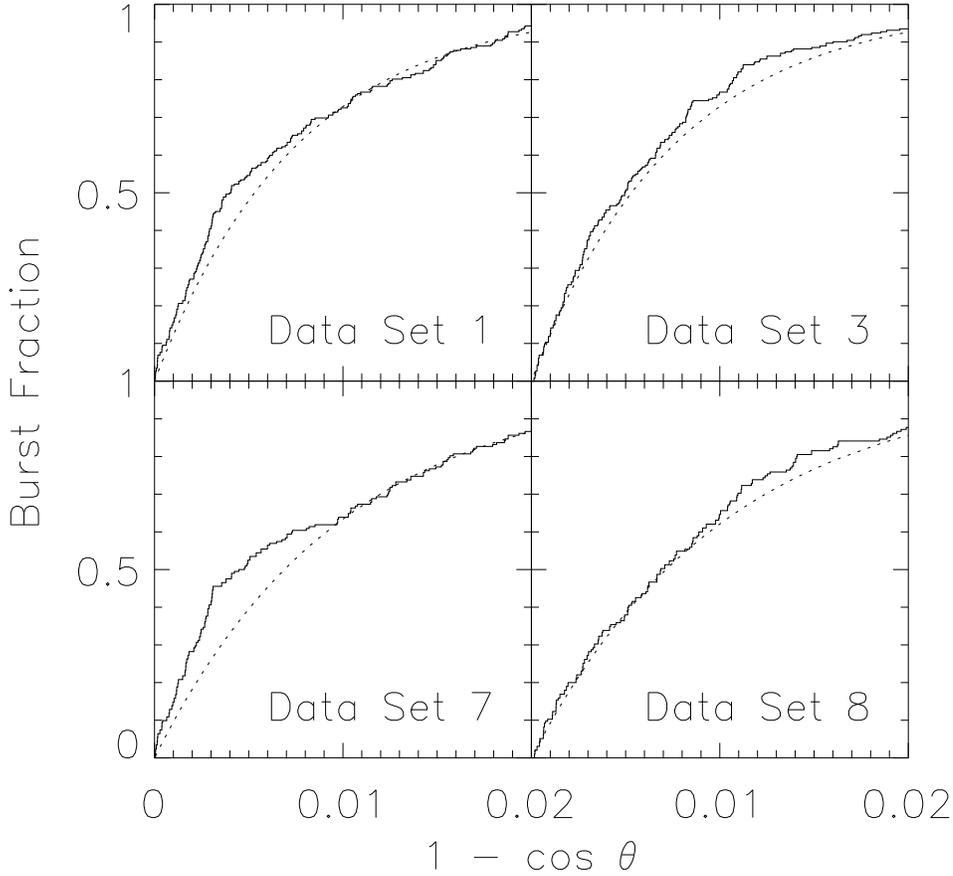

Fig. 2—Nearest neighbor cumulative distributions from the 2B catalog, plotted as functions of $1-\cos\theta$, where $\theta$ is the angle to the nearest neighbor. The data are plotted as a histogram while the model curve for isotropy is plotted as a smooth curve. The four plots are for data sets 1, 3, 7, and 8 of Table 1. Data sets 1 and 7 correspond to the 1B catalog while data sets 3 and 8 correspond to the 2B − 1B catalog. Data sets 1 and 3 are consecutive sets of 262 gamma-ray bursts. Data sets 7 and 8 are bursts with position errors $< 9°$.

Results of our nearest neighbor analysis of the Second BATSE Catalog and various subsets are given in Table 1. Figure 2 shows the nearest neighbor cumulative distribution for four of these subsets. We find the maximum deviation $D$ of each data set from the isotropic cumulative distribution and derive the KS statistic $K = D\sqrt{N_B}$, where $N_B$ is the sample size. The significance $S$ of the magnitude of $K$—that is, the fraction of trials that produce a greater deviation from the model curve—is determined through Monte Carlo simulation; the usual analytic formula is invalid for this analysis because the nearest neighbors are not statistically independent. The results are given in Table 1 for both the celestial coordinate frame ($K_{\text{cel}}$ and $S_{\text{cel}}$) and the CGRO coordinate frame ($K_{\text{gro}}$ and $S_{\text{gro}}$). The analysis in CGRO coordinates is particularly sensitive to systematic effects relating to the angular response of the BATSE detectors. Such effects would be less apparent in the celestial coordinate frame because the CGRO orientation is routinely changed at one or two week intervals. The effect seen by Quashnock & Lamb is reproduced in our data set 7, while the remaining subsets exhibit no statistically significant deviation from isotropy.

An upper limit on the number of repeating sources can be found from both the nearest neighbor test and the farthest neighbor test. Through Monte Carlo simulation we derived these limits for an isotropic distribution of burst sources. The model that the various data sets were tested against consists of $n_1$ sources that each produce one observed burst and $n_r$ sources that each produce $\nu$ observed bursts. The burst locations of the repeating sources are given a Gaussian distribution with a 9° standard deviation about the source location. Limits on the burst repeater fraction are derived from the maximum deviation of the various data sets from the average nearest and farthest neighbor cumulative distributions of the Monte Carlo simulations. Because of the computational demands of this analysis, we determine the significance of a deviation from the probability of a similar deviation in the no-repeater model. We have verified that this approximation is adequate by deriving some significances using simulations of repeater models. The upper limit on the repeater fraction for all sets is roughly the average value of the upper limit found for each data set, which is the smaller of the limits derived from the nearest and the farthest neighbor statistic. We find from the Monte Carlo simulations that the nearest and farthest neighbor statistics limit the quantity $f(\nu - 1)^{1.2}/\nu$. Except for data set 7, we find the limit to be $f(\nu - 1)^{1.2}/\nu \lesssim 0.2$. For $\nu = 2$ the limit is $f \lesssim 0.4$, while for $\nu = 5$ it is $f \lesssim 0.2$.

We test data set 8, which was selected with the 9° criterion of Quashnock & Lamb, against the null hypothesis and find no evidence of burst repetition. As explained below, we calculate that BATSE remains sensitive to repeaters, so the results of data sets 7 and 8 are contradictory. Considering the difficulty of evaluating the significance of the evidence for repeaters found by retrospective analysis of the 1B catalog and the non-confirmation of the effect in the post-1B data, we conclude that the data are consistent with the null hypothesis of isotropy.

## 4. DISCUSSION

If the MAXBC-located bursts are removed from the 2B catalog, the effective exposure decreases. Exposure is defined here as the fraction of bursts above trigger threshold that would be observed, and is less than unity due to earth blockage, SAA passes, and intervals during which the burst trigger is disabled. Our investigation of the effect of exposure on the detectability of repeaters finds that if the number of detectable bursts from a repeating source is large, then the average number of bursts observed for each source remains large, and the fraction of bursts identified as bursts from repeater sources changes little, but if the number of bursts from a repeater source is small, so that the average observed number of bursts is small, then the number of repeater sources that appear as single burst sources increases significantly as the efficiency of detecting a burst drops. The effect of this is to decrease the fraction of bursts that are identifiable as bursts from repeaters.

Figure 3 illustrates the effect of a varying exposure on the detectability of repetitions in one specific model—an ensemble of sources that each produce 10 bursts above threshold, not all of which are detected. The solid line, referenced to the left axis, is the fraction of events that appear to have at least one companion burst. Our simulations have shown that the strength of the signal in the nearest neighbor test is approximately proportional to this curve. The dashed line, referenced to the right axis, is the average number of bursts observed from sources that produce at least two observed bursts. The strength of the signal in the two point correlation function is approximately proportional to this curve. The vertical dashed line at 0.34 indicates the exposure of the 1B catalog. Here, an average of about 3.5 bursts will be observed from each repeating source, and about 3% of the observed bursts will be misidentified as non-repeaters. Note that this number of observed repetitions per observed repeater is slightly less than the value $\nu \gtrsim 4$ suggested by Quashnock & Lamb (1993), based on the angular scale of the clumpings seen in the 1B catalog. A smaller source repetition rate than our choice of 10 produces a smaller observed repetition rate and a larger fraction of misidentifications. The vertical dashed line at 0.25 indicates the exposure of the post-1B portion of the 2B catalog when MAXBC-located bursts have been removed. Here, an average of about 3.0 bursts will be observed from each repeating source, and about 8% of the observed bursts will be misidentified as non-repeaters. For this specific model, a change in exposure of only 26% has a negligible effect on the burst repetition limit derived for the 2B catalog from the nearest neighbor analysis. The limit from the two point angular correlation function is increased by $\sim 15\%$.

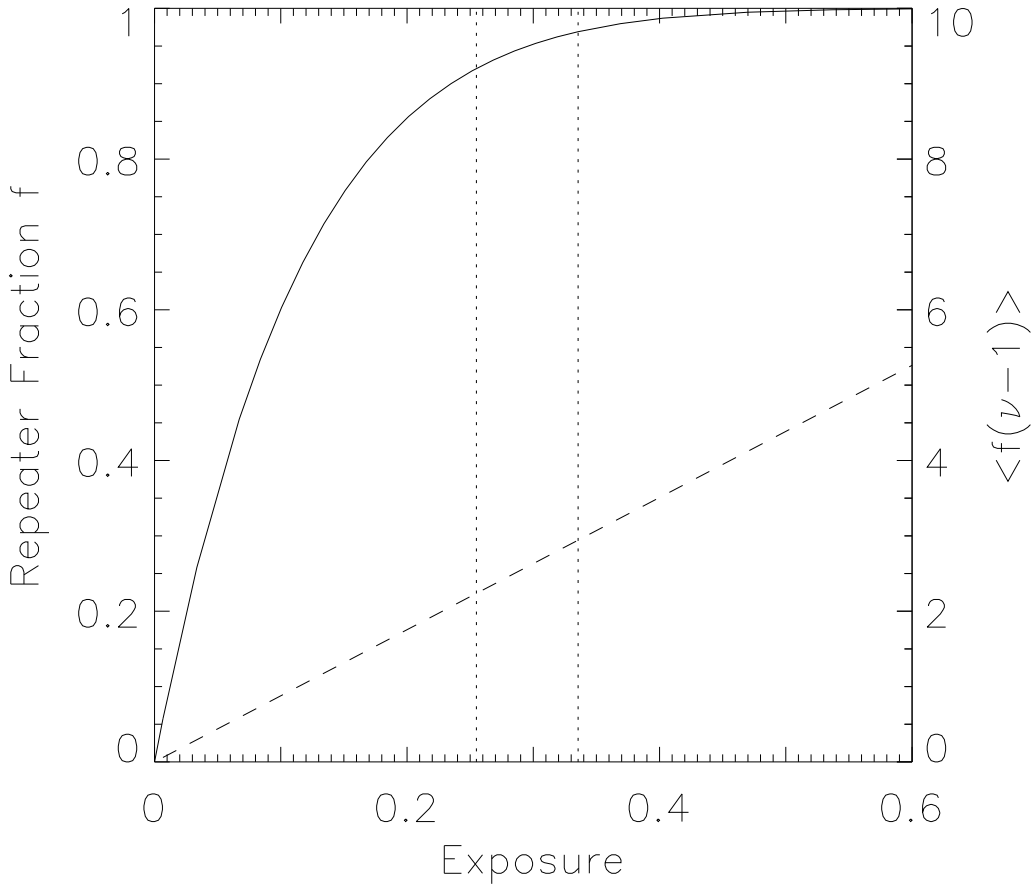

Fig. 3—Efficiency of observing burst repetition as a function of sky exposure, for a model in which each source produces 10 outbursts. The solid line, referenced to the left axis, is the fraction of events that can be identified as repeaters. The dashed line is the average number of bursts observed from sources that produce at least two observed bursts. The right hand vertical line is the exposure for the 1B catalog and the left hand vertical line is the exposure for the 2B − 1B catalog.

While we find the data to be consistent with no burst repetition, repetition at some level cannot be excluded. Using a simple model, we place upper-limits on repetition using the two-point angular correlation function and the nearest and farthest neighbor statistics. In all cases the two-point angular correlation function places a tighter limit on the fraction $f$ of observed bursts that could be from repeating sources than the nearest neighbor statistic. The limit for $\nu = 2$ observed bursts per repeating source is $f \lesssim 0.2$ and for $\nu = 5$ the limit is $f \lesssim 0.05$. We do not place a limit on the fraction of sources that emit multiple bursts because such limits are highly model dependent.

Several factors will improve these results. Statistical limits will be reduced as BATSE continues to accumulate burst locations. Flight software changes since the 2B catalog have eliminated the need for MAXBC locations. The current daily exposure exceeds that of the 1B era, primarily because of reduced solar activity. Finally, we continue to refine the burst location algorithm to reduce systematic errors. As a consequence, analysis of the forthcoming 3B catalog will improve the constraints on burst repetition.

*Acknowledgements:* This work is supported in part by NASA grant NAG 5-1578.